\documentclass[10pt]{article}
\usepackage{hyperref,graphicx}   

\usepackage{multirow}
\usepackage{stackengine}
\usepackage{amsfonts}
\usepackage{amssymb}
\usepackage{amsbsy}
\usepackage{amsmath}
\usepackage{latexsym}
\usepackage{bm}
\usepackage{color}
\usepackage{wasysym}
\usepackage{mathbbol}
\begin{document}
%%%%%%%%%%%%%%%%%%%%
\title{{\bf{\Large Are multiple reflecting boundaries capable of enhancing  entanglement harvesting?}}}
%%%%%%%%%%%%%%%%%%%%
\author{
 {\bf {\normalsize Dipankar Barman}$
$\thanks{E-mail: dipankar1998@iitg.ac.in}},\, 
 {\bf {\normalsize Bibhas Ranjan Majhi}$
$\thanks{E-mail: bibhas.majhi@iitg.ac.in}}\\
 {\normalsize Department of Physics, Indian Institute of Technology Guwahati,}
\\{\normalsize Guwahati 781039, Assam, India}
\\[0.3cm]
}
%\date{}

\maketitle

\begin{abstract}
Quantum entanglement harvesting in the relativistic setup attracted a lot of attention in recent times. Acquiring more entanglement within two qubits may be very desirable to establish fruitful communication between them. On the other hand use of reflecting boundaries in a spacetime has close resemblance to the cavity quantum optomechanical systems. Here, in presence of two reflecting boundaries, we study the generation of entanglement between two uniformly accelerated Unruh-DeWitt detectors which are interacting with the background scalar fields. Like no boundary and single boundary situations, entanglement harvesting is possible for their motions in opposite Rindler wedges. We observe that the reflecting boundaries can play double roles. In some parameter space it causes suppression, while in other parameter space we can have enhancement of entanglement compared to no boundary and single boundary cases. Thus increase of boundaries has significant impact in this phenomena and a suitable choices of parameters provides desirable increment of it.  
\end{abstract}

%%%%%%%%%%%%%%%%%%%%%%%%%%%%%%%%%%%%%%%%%%%%%%%%%%%%%%
\section{Introduction}

Quantum entanglement is one of the predictions of quantum theory, which shows fascinating nonlocal properties. Two observers can be entangled, even if they are spacelike separated. This phenomenon is crucial to many quantum information theoretic processes, such as quantum teleportation  \cite{Hotta:2008uk, Hotta:2009, Frey:2014, Matson:2012aa}, cryptography \cite{PhysRevLett.67.661,Yin:2020aa}, and computation \cite{Mooney:2019aa}.  In recent decades, many studies have been done to understand this phenomenon in relativistic set-up in flat and curved spacetimes.
%In recent decades there has been growing interest in understanding this phenomenon in relativistic set up in flat and curved spacetimes.
 This phenomenon also plays a vital role in understanding the black hole information paradox \cite{hawking1975, Almheiri:2012rt, Marolf_2017}, the quantum nature of gravity \cite{PhysRevLett.119.240401,PhysRevLett.119.240402}, black hole thermodynamics \cite{Brustein:2005vx,Solodukhin:2011gn}, etc. 

Any quantum field theory's vacuum is an entangled state from a local observer's point of view \cite{RevModPhys.76.93}. The Bell-CHSH inequality is maximally violated in the field's vacuum \cite{SUMMERS1985257, doi:10.1063/1.527733, 10.1063/1.527734, Summers:1987ze}. Two Unruh-DeWitt (UDW) detectors \cite{Unruh:1976db,book:Birrell} locally interacting with the background quantum field can extract this entanglement from the quantum vacuum. This process of entanglement extraction is popularly known as `entanglement harvesting'. Entanglement harvesting is possible even if the detectors are causally disconnected and it is independent of the internal structure of the detectors. The formulation for understanding this harvesting phenomenon is first solidified by Reznik \cite{Reznik:2002fz,Reznik:2003mnx} and then further improved in \cite{Salton:2014jaa, Koga:2018the,Ng:2018ilp}, where proper time-ordering is introduced into the picture. These studies usually deal with two two-level detectors interacting with the background field and are in an initial uncorrelated state. To study the characteristics of the harvested entanglement for these bipartite systems, negativity and concurrence are well established as a measure of entanglement. The nature of the harvested entanglement depends on the background geometry \cite{Henderson:2017yuv, Tjoa:2020eqh, Cong:2020nec, Gallock-Yoshimura:2021yok,VerSteeg:2007xs, Henderson:2018lcy,Barman:2021kwg}, boundary conditions \cite{Cong:2018vqx, Cong:2020nec}, motion of the detectors  \cite{FuentesSchuller:2004xp, Martin-Martinez:2015qwa,Salton:2014jaa, Koga:2018the, Koga:2019fqh,Barman:2021bbw,Barman:2021kwg,Barman:2022xht,Barman:2022loh}, etc.

The entanglement dynamics between two detectors is an observer-dependent phenomenon \cite{PhysRevA.55.72, PhysRevLett.88.230402, PhysRevLett.94.078901, PhysRevLett.89.270402, PhysRevLett.91.180404, FuentesSchuller:2004xp, Chowdhury:2021ieg}. It is well known that acceleration of the detectors promotes entanglement between the detectors under certain circumstances  \cite{Reznik:2002fz,Salton:2014jaa,Koga:2019fqh,Barman:2021bbw,Liu:2021dnl}. One can consider the interaction between the detectors and the background field to be eternal, which allows one to avoid the switching effects due to switching functions \cite{book:Birrell}. For eternal interaction in free Minkowski space, it is known that two accelerated detectors can get entangled if they are only in anti-parallel motion. In a recent study \cite{Liu:2020jaj}, the influence of a reflecting boundary on entanglement harvesting between two UDW detectors has studied with finite-time interaction switching function, where it is observed that entanglement between two UDW detectors gets suppressed in the presence of a reflecting boundary. However, there exists a parameter space where a reflecting boundary can enhance the correlation between the detectors. One also observed enhancement in entanglement harvesting near an extremal black hole compared to a non-extremal black hole in some parameter spaces \cite{Barman:2023rhd}. Understanding the role of reflecting boundaries is crucial due to its applicability to cavity quantum optomechanical systems (cavity QED) with numerous practical applications \cite{book:CQED2006}. Reflecting boundaries also plays an essential role in the context of holographic entanglement entropy \cite{Akal:2021foz}, secure quantum communication over long distances \cite{2019QuIP...18...37H, 2020EPJD...74..176H, Good:2021asq}, the Casimir-Polder interaction \cite{PhysRevA.76.032107, PhysRevA.76.062114, PhysRevA.82.042108}, the radiative properties of atoms  \cite{Scully:2003zz, Yu:2005ad, Yu:2006kp, Belyanin2006, Rizzuto_2009, Arias:2015moa,sym11121515,  riddhi21}, the geometric
phase \cite{ZHAI2016338} and the modified entanglement dynamics \cite{Zhang:2007ha,Cheng:2018nhg}, etc.

As we mentioned in the earlier work \cite{Liu:2020jaj}, presence of a reflecting boundary can suppress or enhance entanglement harvesting between two detectors, depending on the parameter space under consideration. One can ask whether the similar effect of boundary also holds in presence of multiple boundaries. Will these effects -- entanglement suppression and enhancement due to the presence of a reflecting boundary be amplified if multiple reflecting boundaries are present there? It will be much interesting if one finds more enhancement in the harvested entanglement in presence of multiple boundaries. Till now, there has been no studies done so far, analysing the entanglement harvesting phenomena in presence of multiple reflecting boundaries. In this study, we have done a detailed analysis of such a phenomena, considering two reflecting boundaries. We compare the harvested entanglement in presence of double boundaries with the entanglement harvested in single and no boundary systems. Here we consider the reflecting boundaries are extended in the $x$-$y$ plane and located at $z=0$ and $z=L$. The detectors accelerate along the $x$-direction.
%{\color{red}  If one sets $L=0$ or $\infty$, the system will reduce to a single boundary system.} 
%And, then taking the detectors far away $z=0$ plane, one should achieve the results for no boundary system.    \\Since the presence of boundaries modify the vacuum fluctuations of the local quantum field, the Wightman functions will be modified. 
We use Green's functions for two reflecting boundaries as provided in \cite{PhysRev.184.1272, book:Birrell, riddhi21} and follow the formulation for entanglement harvesting utilized in \cite{Koga:2018the, Koga:2019fqh}. To study the fate of entanglement between two detectors, we investigate the concurrence \cite{Bennett:1996gf, Hill:1997pfa, Wootters:1997id}, as a measure of the harvested entanglement. 
% In particular, we consider the interaction between two two-level detectors and the scalar field to be of monopole type.  Similar to \cite{Koga:2019fqh, Barman:2021bbw}, we also assert that entanglement extraction is possible only for the anti-parallel motion of the detectors and not for the parallelly accelerated ones for eternal switching of the interaction. We also encounter the phenomena of degradation entanglement extraction with increasing separation between the detectors. 
Here we consider three types of arrangements for detector trajectories to apprehend the effect of the boundaries.  
First, we take one detector near the first boundary and another near the second boundary, equally distanced from $z=L/2$. 
Second, we take both the detectors in the same $z$ positions, $i.e.,$ $z_A=z_B$. 
Third, we fixed the position of one detector in-between regions of the boundary planes and moved another detector to understand the influence of the boundaries. 
We observe that entanglement enhancement and suppression are also possible is presence of double boundaries. The enhancement and suppression of harvested entanglement due to the presence of boundaries is more perceptible for the double boundary system.

This paper is organized as follows. In Sec. \ref{sec:2}, we 
discuss the framework for Entanglement harvesting between two  UDW detectors interacting with a minimally coupled,
massless scalar field through monopole terms. This section
discusses the mathematical description of the entanglement harvesting condition. In Sec. \ref{sec:3}, we discuss the trajectories and the green functions of the two accelerated UDW
detectors in the presence of reflecting boundaries and investigate the individual detector transition probabilities. Subsequently, in Sec. \ref{sec:4}, we discuss the possibility of entanglement harvesting between the detectors in parallel and anti-parallel motion. Also, the properties of the harvested entanglement are being analysed. Finally, in Sec. \ref{sec:5}, we conclude with the overall discussion of the results.

%%%%%%%%%%%%%%%%%%%%%%%%%%%%%%%%%%%%%%%%%%%%%%%%%%

\section{The model: framework for entanglement harvesting
}\label{sec:2}
We now briefly present our model of two UDW detectors which are simultaneously interacting with background massless real scalar fields. Following the analysis of \cite{Koga:2018the,Ng:2018ilp} the main working formulas, valid till the second order perturbative expansion, will be given. 
%This section presents the formulation for understanding the possibility of two uncorrelated atomic 
%Unruh-DeWitt detectors getting entangled over time while interacting with a general field state. Particularly we are interested to find out the condition for the detectors to be entangled and also the quantification of it will be done. The whole analysis will be valid till the second order perturbative series of the density matrix of the system, when the expansion is done order by order in terms of interaction strength.

Let us consider two observers, Alice and Bob, with two-level Unruh-DeWitt detectors, denoted 
as $A$ and $B$. We consider the detectors point-like and interacting with a massless, real scalar field $\phi(X)$ through monopole interaction. 
Then the interaction action is given by
\begin{eqnarray}
S_{i n t} &=&\sum_{j=A,B}\lambda_{j} \int_{-\infty}^{\infty}  d \tau_{j}\,\kappa_{j}\left(\tau_{j}\right) 
m_{j}\left(\tau_{j}\right) \phi\left(x_{j}\left(\tau_{j}\right)\right)\,,
\end{eqnarray}
where $\lambda_{j}$ is the coupling constant between the $j^{th}$ detector ($j=A,B$) and the scalar field, $\kappa_{j}\left(\tau_{j}\right)$ and $\tau_{j}$ are the interaction switching function and proper time for the $j^{th}$ detector, respectively. 
The monopole operator of the detector's are taken as 
\begin{equation}\label{M}
m_{j}(\tau_{j})=e^{iH_{j}\tau_{j}}(|e_{j}\rangle\langle g_{j}|+|g_{j}\rangle\langle e_{j}|)e^{-iH_{j}\tau_{j}}\,.
\end{equation}
 Here $|g_{j}\rangle$ and $|e_{j}\rangle$ are the ground and exited states of the $j^{th}$ detector, respectively.

The initial state of the composite system is taken to be 
$|in\rangle=|0_{M}\rangle|E_{0}^{A}\rangle|E_{0}^{B}\rangle$, where $|0_{M}\rangle$ is the Minkowski vacuum, the state of the field and $|E_{n}^{j}\rangle~(n=0,1)$ is the $n^{th}$ state of the $j^{th}$ detector. The final state of the system in the asymptotic future can be obtained as $|out\rangle=T\{e^{iS_{int}}\}|in\rangle$. 
One can get the reduced density matrix for the detectors $\rho_{AB}$ by tracing out the field degrees of freedom from the final total density matrix, which in the basis of 
$\{|E_{1}^{A}\rangle|E_{1}^{B} \rangle,|E_{1}^{A} \rangle|E_{0}^{B}\rangle, 
|E_{0}^{A}\rangle|E_{1}^{B} \rangle,|E_{0}^{A}\rangle|E_{0}^{B}\rangle\}$ is expressed as \cite{Koga:2018the}
%
%%%%%%%%%%%%%%%%%%%%%%%%%%%%%%%%%%%%%%%%%%%%%%%%%%%%%%%%%%%%%%%%%
\begin{equation}\label{eq:detector-density-matrix}
\rho_{AB}=\begin{pmatrix}
	0			&0					&0				&\lambda^{2}\mathcal{E}\\
	0			&\lambda^{2}\mathcal{P}_{A}		&\lambda{2}\mathcal{P}_{AB}	&0\\
	0			&\lambda^{2}\mathcal{P}^{\star}_{AB}	&\lambda^{2}\mathcal{P}_{B}	&0\\	
\lambda^{2}\mathcal{E}^{\star}	&0					&0				&1-\lambda^{2}\mathcal{P}_{A}-\lambda^{2}\mathcal{P}_{B}
\end{pmatrix}+O(\lambda^{4}).
\end{equation}
Here for simplification we choose $\lambda_{A}=\lambda_{B}=\lambda$. 
 The structure of the density matrix depends on the choice of the initial detectors' state and the monopole operator considered. For our particular monopole operator given in Eq. (\ref{M}), expressions for the detectors' density matrix elements are
\begin{eqnarray}\label{IExpressions}
\mathcal{P}_{j}&=&\int_{-\infty}^{\infty}\int_{-\infty}^{\infty}{d\tau_{j}d\tau'_{j}}\,\kappa_j(\tau_j)\kappa_j(\tau'_j)e^{
-i\Delta{E}(\tau_{j}-\tau'_{j})}\,G_{W}(x_{j},x'_{j})\,,
%\nonumber\\&&~~~~~~~~~~~~~~~~~~~~~~~~~~\times\langle\Psi|\phi(x_{j})\phi(x'_{j})|\Psi\rangle\,,
\nonumber\\ 
\mathcal{E}&=&-\int_{-\infty}^{\infty}\int_{-\infty}^{\infty}{d\tau_{B}d\tau'_{A}}\,\kappa_B(\tau_B)\kappa(\tau'_A)
e^{i\Delta{E}(\tau'_{A}+\tau_{B})}\,iG_{F}(x_{B},x'_{A})\,,
%\nonumber\\&&~~~~~~~~~~~~~~~~~~~~~~~\times\langle\Psi|T\phi(x_B)\phi(x'_A)|\Psi\rangle\,,
\nonumber\\
\mathcal{P}_{AB}&=&\int_{-\infty}^{\infty}\int_{-\infty}^{\infty}{d\tau_{B}d\tau'_{A}}\,\kappa_B(\tau_B)\kappa_A(\tau'_A)e^{i\Delta{E}(\tau'_{A
}-\tau_{B})}\,G_{W}(x_{B},x'_{A})
%\nonumber\\&&~~~~~~~~~~~~~~~~~~~~~~~~~~\times \langle\Psi|\phi(x_{B})\phi(x'_{A})|\Psi\rangle
\,.
 \end{eqnarray}
%
%\textcolor{red}{Here the contributions from the monopole operators are appeared in the exponentials}. 
The quantities $G_{W}(x_{i},x'_{j}),\,G_{F}(x_{i},x'_{j})$ are respectively the positive frequency Wightman function and the Feynman propagator, defined as
\begin{eqnarray}
G_{W}(x_{i},x'_{j})&=&\langle0_{M}|\phi(x_{i})\phi(x_{j})|0_{M}\rangle\,,\nonumber\\
iG_{F}(x_{i},x'_{j})&=&\langle0_{M}|T\{\phi(x_{i})\phi(x_{j})\}|0_{M}\rangle
\,~.
\end{eqnarray}
%
%\textcolor{red}{These contain the information about the detectors motion and the background spacetime with mirrors.}
The detailed analysis of the density matrix one may look into \cite{Koga:2018the}. The detectors are taken to be identical and hence we denoted $\Delta E = E^j_1-E^j_0$ for all $j$.%

Since our system is a bipartite system, any negative  eigenvalue of the partial transposition of the reduced density matrix (see Eq.(\ref{eq:detector-density-matrix})) confirms entanglement between the detectors \cite{Peres:1996dw, Horodecki:1996nc}. The absolute value of sum of all negative eigenvalues is known as {the {negativity}}, a measure of entanglement. For our density matrix, there will be a negative eigenvalue if the following condition is satisfied \cite{Koga:2018the, Koga:2019fqh}
\begin{equation}\label{eq:cond-entanglement}
 \mathcal{P}_{A}\,\mathcal{P}_{B}<|\mathcal{E}|^2~.
\end{equation}%
Once the above condition is satisfied, one may 
study various measures to quantify the harvested entanglement. In this regard, a convenient entanglement measures is the concurrence (defined as $\mathcal{C}(\rho_{AB})= {\it max}\{0,\,2\lambda^2\mathcal{C}_{J}(\rho_{AB})$\}) \cite{Bennett:1996gf, Hill:1997pfa, Wootters:1997id, Koga:2018the}, which is very useful for estimating the entanglement 
of formation $E_{F}(\rho_{AB})$ (see \cite{Bennett:1996gf, Hill:1997pfa, Wootters:1997id, Koga:2018the}). For our two qubits system the quantity $\mathcal{C}_{J}$ is obtained as  \cite{Koga:2018the} 
\begin{eqnarray}\label{eq:concurrence-gen-exp}
&&\mathcal{C}_{J}(\rho_{AB}) = 
 \left(|\mathcal{E}|-\sqrt{\mathcal{P}_{A}\mathcal{P}_{B}}\right)~.
\end{eqnarray}
This quantity $\mathcal{C}_{J}$ can have both positive or negative values. Due to the definition of the concurrence ($\mathcal{C}$), negative values of $\mathcal{C}_{J}$ implies $zero$ concurrence of the two detector system. Therefore, harvesting entanglement between the detectors required positivity of the quantity $\mathcal{C}_{J}$. Also note that positivity of this quantity automatically implies the condition (\ref{eq:cond-entanglement}). 
Therefore, studying this quantity enables us to understand the characteristics of entanglement between two detectors due to the background spacetime and motions of detectors with specific configurations. For simplicity of the model and analytically handling the computations, like in earlier investigations \cite{Koga:2018the,Koga:2019fqh,Barman:2021bbw,Barman:2021kwg,Barman:2022xht,Barman:2022loh}, we will consider eternal interaction; i.e. $\kappa_j =1$ in the subsequent analysis. It may be pointed that $\mathcal{P}_j$ can be identified as the individual detector's transition probability, whereas in literature $\mathcal{E}$ is usually called as entangling term. 
%%%%%%%%%%%%%%%%%%%%%%%%%%%%%%%%%%%%%%%%%%%%%%%%%%%%%%%%%%%%%%%%%%%%%

\section{Accelerated detectors with reflecting boundaries}\label{sec:3}

Let us consider the $(3+1)$-dimensional Minkowski spacetime (coordinates are denoted as ($t,x,y,z$)) with two parallel reflecting boundaries extended in the $x$-$y$ plane -- one is at $z=0$ and another at $z=L$. In the context of cavity quantum electrodynamics, the quantity $L$ is known as cavity length. The positive frequency Wightman function for a massless scalar field in $(3+1)$-dimensional Minkowski spacetime in the presence of reflecting boundaries is given by \cite{PhysRev.184.1272, book:Birrell, riddhi21}\footnote{In the book by Birrell and Davies \cite{book:Birrell}, the factor of $2$ with $L$ in the green function of Eq.(\ref{green+2}) is missing (we feel it is a typo). The same factor can be found in the original work  \cite{PhysRev.184.1272}, and in a recent work \cite{riddhi21}.}

\begin{eqnarray}\label{green+2}
G_{W}(x,x')&=&-\frac{1}{4\pi^{2}}\sum_{n=-\infty}^{\infty}\left(\frac{1}{(t-t'-i\epsilon)^{2}-(x-x')^{2}-(y-y')^{2}-(z-z'-2L\,n)^{2}}\right.\nonumber\\&&~~~~~~~-\left.\frac{1}{(t-t'-i\epsilon)^{2}-(x-x')^{2}-(y-y')^{2}-(z+z'-2L\,n)^{2}}\right)\,.
\end{eqnarray}
By construction, this green function vanishes at $z$ (or $z')=0$ or $L$. Considering only $n=0$ term, one gets the Wightman function in presence of a single reflecting boundary at $z=0$. In this particular situation, among two terms -- the first term corresponds to the unbounded Minkowski space and the second term is due to the boundary effect. 

The trajectories of the detectors uniformly accelerating along $x$-direction in terms of their proper times are given by \cite{book:Birrell, Koga:2019fqh, Barman:2021bbw}
\begin{eqnarray}\label{Trajec}
&&t_{A}=a_{A}^{-1}\sinh(a_{A}\tau_{A}),~x_{A}=a_{A}^{-1}\cosh(a_{A}\tau_{A}),~y_{A}=0,~z_{A}=z_{A}\,;
\nonumber\\
&&t_{B}=a_{B}^{-1}\sinh(a_{B}\tau_{B}),~x_{B}=\pm a_{B}^{-1}\cosh(a_{B}\tau_{B}),~y_{B}=\Delta{y},~z_{B}=z_{B}\,~,
\end{eqnarray}
where $0< z_{A},\,z_{B}<L$. $a_{A}$ and $a_{B}$ are respectively the acceleration of the detectors $A$ and $B$. The ``$+$'' ($-$) sign in $x_{B}$ corresponds to motion of the detector $B$ in the right (left) Rindler wedge. 

Now, let us calculate $\mathcal{P}_j$. The denominators of the Wightman function in (\ref{green+2}) for a single detector (i.e. any one of the detectors among $A$ or $B$ is moving either in left or in right Rindler wedge) are evaluated below. Here we can drop the detector subscripts as these quantities are same for any detector. Using the trajectories (\ref{Trajec}), one obtain the denominators of the first and second terms in the parenthesis of (\ref{green+2}) as
\begin{eqnarray}\label{Deno}
&&(t-t'-i\epsilon)^{2}-(x-x')^{2}-(y-y')^{2}-(z-z'-2L\,n)^{2}=
4a^{-2}(\sinh^{2}(a(\tau-\tau')/2-i\epsilon)-a^{2}L^{2}n^{2})\,,\nonumber\\
&&(t-t'-i\epsilon)^{2}-(x-x')^{2}-(y-y')^{2}-(z+z'-2L\,n)^{2}=
4a^{-2}(\sinh^{2}(a(\tau-\tau')/2-i\epsilon)\nonumber\\&&~~~~~~~~~~~~~~~~~~~~~~~~~~~~~~~~~~~~~~~~~~~~~~~~~~~~~~~~~~~~~~~~~~~~~~-a^{2}({z}-L\,n)^{2})\,.
\end{eqnarray}
These two quantities have the same proper time dependence with different additional constants. Hence one can write them in a combined way as $4a^{-2}(\sinh^{2}(a(\tau-\tau')/2-i\epsilon)-g_{n}^{2})$ with $g_{n}$ as $L\,a\,n$ and $a(z+L\,n)$ for the first and second denominators, respectively. To calculate the transition probability, we need to perform time-integrations in the first equation of (\ref{IExpressions}) by using (\ref{green+2}) and (\ref{Deno}). The two terms in the Wightman function provide identical integrations and therefore performing the following form of integration is sufficient to achieve the goal. Following \cite{Koga:2019fqh} one finds
\begin{equation}
-\frac{a^{2}}{16\pi^{2}}\int\int{d\tau d\tau'}\frac{e^{i\Delta{E}(\tau-\tau')}}{\sinh^{2}(a(\tau-\tau')/2+i\epsilon)-g_{n}^{2}}=\frac{\delta(0)}{2\left(e^{\pi  \alpha }-1\right) }\frac{\sin \left(\alpha  \sinh ^{-1}(|g_{n}|)\right)}{|g_{n}| \sqrt{g_{n}^2+1}}\,,
\end{equation}
where, $\alpha=2\Delta E/a$. % and the quantity $g$ have values $A\,a\,n$ and $a(A\,n\,+\,\Delta z)$ for two terms in the parenthesis of the green function given in (\ref{green+2}). 
Here we used coordinate transform $T=(\tau+\tau')/2,~\sigma=\tau-\tau'$ and performed contour integral over $\sigma$-variable. Then the transition probability of $j^{th}$-detector is obtained as
\begin{equation}\label{Pj}
\begin{aligned}
\mathcal{P}_{j}&=\int\int{d\tau d\tau'}e^{i\Delta{E}(\tau-\tau')}G_{W}(x',x)\\&=
\frac{\delta(0)}{2\left(e^{\pi  \alpha }-1\right) }\sum_{n=-\infty}^{\infty}\left(\frac{\sin \left(\alpha  \sinh ^{-1}(|L\, a\, n|)\right)}{|L\, a\, n| \sqrt{|L\, a\, n|^2+1}}-
\frac{\sin \left(\alpha  \sinh ^{-1}(|a(z_{j}+L\, n)|)\right)}{|a(z_{j}+L\, n)| \sqrt{|a(z_{j}+L\, n)|^2+1}}
\right)\,.
\end{aligned}
\end{equation}
This will be needed for testing the validity of the entangling condition (\ref{eq:cond-entanglement}) and the calculation of concurrence (\ref{eq:concurrence-gen-exp}).

Before proceeding to this, few comments are in order.
First of all, note that when $Ln/z_j>>1$, two term in (\ref{Pj}) will cancel each other. This can happen for large values of $n$ and therefore we may set a cut off on upper and lower limit of $n$ in order to evaluate the summation in (\ref{Pj}). Therefore later in numerical calculation we choose $max |n|$ to a large finite value. Secondly, only $n=0$ term in (\ref{Pj}) refers to the transition probability of an accelerated detector with single reflecting boundary. Finally, the term for $n=0$ in the first part reproduces the same in unbounded Minkowski spacetime (see, \cite{Koga:2019fqh}):
\begin{equation}\label{Pj1}
\begin{aligned}
P_{j}&=
\frac{\delta(0)\Delta E}{a\,\left(e^{\pi  \alpha }-1\right) }\,.
\end{aligned}
\end{equation}

%%%%%%%%%%%%%%%%%%%%%%%%%%%%%%%%%%%%%%%%%%%%%%%%%%%%%%%%%%%%%%%%%%%%%
\section{Entanglement harvesting}\label{sec:4}
In this section, we will evaluate the entangling term for the accelerated detectors. First we will calculate it for the parallel motion of the detectors; i.e. the detectors are in same Rindler wedge, namely in right wedge. Then the anti-parallel motion; i.e. one detector is in right wedge and other one is in left wedge, will be considered.

\subsection{Parallel acceleration}
The evaluation of the quantity $\mathcal{E}$ requires the Feynman propagator, which can be expressed as
\begin{eqnarray}\label{green+2Feyn}
G_{F}(x_{A},x_{B})&=&\frac{i}{4\pi^{2}}\sum_{n=-\infty}^{\infty}\left(\frac{1}{(t_{A}-t_{B})^{2}-(x_{A}-x_{B})^{2}-\rho_{n,-}^{2}-i\epsilon}\right.\nonumber\\&&~~~~~~~-\left.\frac{1}{(t_{A}-t_{B})^{2}-(x_{A}-x_{B})^{2}-\rho_{n,+}^{2}-i\epsilon}\right)\,,
\end{eqnarray}
where $\rho^{2}_{n,\pm}=\Delta y^{2}+(z_{A}\pm z_{B}-2L \,n)^{2}$.  We need to numerically analyse the final outcomes for our later purpose to compare the concurrence quantity for different boundary systems. 
As mentioned earlier, we must have $0<z_{A},\,z_{B}<L$. Now for a finite fixed value of $\Delta{y}$, if we consider $n$ is sufficiently large, then the last term in $\rho_{n,\pm}^{2}$ will dominate. Hence, one will have $\rho_{n,\pm}^{2}\approx4L^{2}n^{2}$. Therefore the quantities inside the parenthesis of Eq. (\ref{green+2Feyn}) corresponding to large $n$ will cancel each other. Thus the infinite summation in Eq. (\ref{green+2Feyn}) effectively can be replaced by a finite summation (the same is also true for the anti-parallel acceleration of the detectors).

The quantities in the denominators can be re-expressed using the detector trajectories given in Eq. (\ref{Trajec}). For parallel motion of the detectors (with '+' sign in $x_{B}$), one obtains
\begin{equation}\label{Ref1}\begin{aligned}
&(t_{A}-t_{B})^{2}-(x_{A}-x_{B})^{2}-\rho_{n,\pm}^{2}-i\epsilon\\
&=\frac{1}{a_{A}a_{B}}\left[e^{a_{A}\tau_{A}-a_{B}\tau_{B}}+e^{-a_{A}\tau_{A}+a_{B}\tau_{B}}-\left(\frac{a_{A}}{a_{B}}+\frac{a_{B}}{a_{A}}+a_{A}a_{B}\rho
_{n,\pm}^{2}\right)\right]-i\epsilon\\
&=\frac{1}{a_{A}a_{B}x}(u-M_{n,\pm}+\sqrt{M_{n,\pm}^{2}-1}+i\epsilon)(u-M_{n,\pm}-\sqrt{M_{n,\pm}^{2}-1}-i\epsilon)\,.
\end{aligned}\end{equation}
Here we define $M_{n,\pm}=\left({a_{A}}/{a_{B}}+{a_{B}}/{a_{A}}+a_{A}a_{B}\rho_{n,\pm}^{2}\right)/2$ and $u=e^{a_{A}\tau_{A}-a_{B}\tau_{B}}$. Performing integration over the variables $\tau_{A}$ and $\tau_{B}$, one obtain the final expression for $\mathcal{E}$ given in (\ref{IExpressions}) as  %The propagator has poles at \begin{equation}\tau_{A}=(a_{B}\tau_{B}\pm\sigma_{\pm}+2\pi i n\pm i\epsilon)/a)_{A};~~(\sigma_{\pm}=\log(M_{\pm}+\sqrt{M_{\pm}^{2}-1}))\end{equation}
\begin{eqnarray}\label{Ref2}
\mathcal{E}(\Delta E)&=&\sum_{n}\frac{i}{2\sqrt{M_{n,-}^{2}-1}}\frac{\delta(\frac{\Delta E}{a_{A}}+\frac{\Delta E}{a_{B}})}{1-e^{-\frac{2\pi\Delta E}{a_{A}}}}\left\{e^{\frac{i\Delta E}{a}\sigma_{n,-}}-e^{-\frac{2\pi\Delta E}{a_{A}}}e^{-\frac{i\Delta E}{a}\sigma_{n,-}}\right\}
\nonumber\\&-&
\frac{i}{2\sqrt{M_{n,+}^{2}-1}}\frac{\delta(\frac{\Delta E}{a_{A}}+\frac{\Delta E}{a_{B}})}{1-e^{-\frac{2\pi\Delta E}{a_{A}}}}\left\{e^{\frac{i\Delta E}{a}\sigma_{n,+}}-e^{-\frac{2\pi\Delta E}{a_{A}}}e^{-\frac{i\Delta E}{a}\sigma_{n,+}}\right\}\,\,,
\end{eqnarray}
where $\sigma_{n,\pm}=\log\left(M_{n,\pm}+\sqrt{M_{n,\pm}^2-1}\right)$. The steps of the above calculation is not new. It has been followed from Sec. III of \cite{Koga:2019fqh}.
The expression in (\ref{Ref2}) contains the Dirac-delta function with argument $\frac{\Delta E}{a_{A}}+\frac{\Delta E}{a_{B}}$, which always vanishes as $\Delta E, \,a_{A},\,a_{B}$ all are positive. On the other hand $\mathcal{P}_j$ is always positive and non-vanishing quantity. Since possibility of entanglement harvesting requires to satisfy the condition (\ref{eq:cond-entanglement}), entanglement harvesting is not possible for parallel motion of the detectors. This situation is similar to the case where no reflecting boundary is considered \cite{Koga:2019fqh, Barman:2021bbw}. The same result is also valid for the identical system with single reflecting boundary as well.

\subsection{Anti-parallel acceleration}
For the anti-parallel motion of the detectors, the trajectories are given in (\ref{Trajec}) with `$-$' sign in the $x_{B}$. Therefore we find
\begin{equation}\label{Ref1}\begin{aligned}
&(t_{A}-t_{B})^{2}-(x_{A}-x_{B})^{2}-\rho_{n,\pm}^{2}-i\epsilon\\
&=-\frac{1}{a_{A}a_{B}}\left[e^{a_{A}\tau_{A}+a_{B}\tau_{B}}+e^{-a_{A}\tau_{A}-a_{B}\tau_{B}}+\left(\frac{a_{A}}{a_{B}}+\frac{a_{B}}{a_{A}}+a_{A}a_{B}\rho
_{n,\pm}^{2}\right)\right]-i\epsilon\\
&=-\frac{1}{a_{A}a_{B}y}(v+M_{n,\pm}+\sqrt{M_{n,\pm}^{2}-1}+i\epsilon)(v+
M_{n,\pm}-\sqrt{M_{n,\pm}^{2}-1}-i\epsilon)
\end{aligned}\end{equation}
where we define $v=e^{a_{A}\tau_{A}+a_{B}\tau_{B}}$. Again, performing integration over
the variable $\tau_{A}$ and then over $\tau_{B}$, one obtain the final expression for $\mathcal{E}$ given in (\ref{IExpressions}) as 
\begin{equation}\label{Ent2}
\begin{aligned}
\mathcal{E}(\Delta E)&=-\frac{1}{2}\frac{\delta\left(\frac{\Delta{E}^{}}{a_{A}}-\frac{\Delta{E}^{}}{a_{B}}\right)}{\sinh\left(\pi\frac{\Delta{E}^{}}{a_{A}}\right)}\sum_{n=-\infty}^{\infty}\left(
\frac{\sin\left(\frac{\Delta{E}^{}\sigma_{n,-}}{a_{A}}\right)}{\sinh(\sigma_{n,-})}-
\frac{\sin\left(\frac{\Delta{E}^{}\sigma_{n,+}}{a_{A}}\right)}{\sinh(\sigma_{n,+})}
\right)~.
\end{aligned}
\end{equation}
Here again, the steps of Sec. III in \cite{Koga:2019fqh} have been followed.

For numerical analysis, practically we do not have to sum over infinite terms. In fact the terms up to $n=N$ (i.e. terms from $-N$ to $N$), where $N$ is chosen to be sufficiently large such that conditions mentioned below Eq. (\ref{green+2Feyn}) are satisfied, will be enough to consider. The reasons are as follows.
First, notice that the entangling term contains the terms like $f(\sigma_{n,-})-f(\sigma_{n,+})$ where $f(\sigma_{n,\pm})=\sin\big(\Delta{E}\sigma_{n,\pm}/a_{A}\big)/\sinh(\sigma_{n,\pm})$. The numerator of this function can have values between $-1$ and $1$. However, the denominator is a massive number for large value of $Ln$. Therefore the larger values of $n$, $f(\sigma_{n,\pm})$ becomes smaller and smaller, and ultimately can be negligible. 
Second, note that the quantities $\sigma_{n,\pm}$ have $n$-dependence through $\rho_{n,\pm}$ (defined below Eq. (\ref{green+2Feyn})). Moreover we already observed that $\rho_{n,+}\approx\rho_{n,-}$ for large value of $Ln$ with $\Delta{y},\,\,z_{A}$ and $z_{B}$ satisfy the earlier mentioned conditions (see the discussion after Eq. (\ref{green+2Feyn})). Thus, a finite summation will be sufficient for the numerical analysis.
We have also verified this feature numerically, where it turns out that the quantity $\mathcal{E}$ becomes constant after a significantly large value of $N$. For instance corresponding to our chosen fixed-parameters, we found that $N=2000$ is enough for our purpose. This is because just below $N=2000$ and above it $\mathcal{E}$ becomes constant (see Fig. \ref{fig:figures5} in the Appendix \ref{AppenA}).

As we mentioned earlier, only the $n=0$ term provides us the effect of single reflecting boundary at $z=0$. Using this we obtain the entangling term in the presence of a single reflecting boundary as

\begin{equation}\label{Ent1}
\begin{aligned}
\mathcal{E}(\Delta E)&=-\frac{1}{2}\frac{\delta\left(\frac{\Delta{E}^{}}{a_{A}}-\frac{\Delta{E}^{}}{a_{B}}\right)}{\sinh\left(\pi\frac{\Delta{E}^{}}{a_{A}}\right)}\left(
\frac{\sin\left(\frac{\Delta{E}^{}\sigma_{0,-}}{a_{A}}\right)}{\sinh(\sigma_{0,-})}-
\frac{\sin\left(\frac{\Delta{E}^{}\sigma_{0,+}}{a_{A}}\right)}{\sinh(\sigma_{0,+})}
\right)~.
\end{aligned}
\end{equation}
Among the two terms of the above, the first term corresponds to the entangling term for the unbounded Minkowski space. Therefore the same with no boundary situation is given by	
\begin{equation}\label{Ent0}
\begin{aligned}
\mathcal{E}(\Delta E)&=-\frac{1}{2}\frac{\delta\left(\frac{\Delta{E}^{}}{a_{A}}-\frac{\Delta{E}^{}}{a_{B}}\right)}{\sinh\left(\pi\frac{\Delta{E}^{}}{a_{A}}\right)}\frac{\sin\left(\frac{\Delta{E}^{}\sigma_{0,-}}{a_{A}}\right)}{\sinh(\sigma_{0,-})}\,~.
\end{aligned}
\end{equation}
%which is independent of distance from the boundaries. However 
Like the entangling terms with single or double boundaries, it also depends on the perpendicular separation between the detectors' trajectories. All of the entangling terms contain the Dirac-delta function with the argument of $\Delta{E}(1/a_{A}-1/a_{B})$. Therefore, to harvest a non-zero amount of entanglement, one must take $a_{A}=a_{B}$. 

Note that when the entangling term is non-vanishing, it contains $\delta(0)$ (like $\mathcal{P}_{j}$ in Eq. (\ref{Pj})). Thus the  quantity ($\mathcal{C}_{J}$) in (\ref{eq:concurrence-gen-exp}) can be expressed as $\mathcal{C}_{J}=\,\delta(0)\mathcal{C}_{I}$, where $\mathcal{C}_{I}$ is a finite quantity. This is a well-known artefact of the choice eternal interaction between the detectors and the background quantum field. However, in order to quantify entanglement through concurrence it is legitimate to define concurrence per unit time, which is given by the positive values of $\mathcal{C}_I$.  This proposal is already well known in literature \cite{book:Birrell, Koga:2019fqh, Barman:2021bbw}. In our later analysis, we only focus on the quantity $\mathcal{C}_{I}$. Now it is time to study $\mathcal{C}_I$ to understand the features of entanglement harvesting. This will be done numerically. For that we introduce dimensionless parameters $\bar{z}_{j}=z_{j}\Delta{E}$, $\bar{\Delta}y=\Delta{y}\Delta{E}$, $\bar{a}_{j}=a_{j}/\Delta{E}$, and $\bar{L}=L_{0}\Delta{E}$. Here we consider, $L_{0}=L$  for the double boundary system (where $L$ is position of the second boundary); otherwise $L_{0}$ is just a numerical parameter, which determines the intra-distance between the detectors.
For our numerical analysis, we choose $\bar{\Delta}y=0.1$ and use solid, dotted and dashed lines to represent no boundary, single boundary and double boundary systems, respectively.

%%%%%%%%%%%%%%%%%%%%%%%%%%%%%%%%%%%%%%%%%%%%%%%%%%%%%%%%%%%%

\subsubsection{Case-I}
\begin{figure}[h!]
	\centering
	\footnotesize
\stackunder[5pt]{\includegraphics[width=0.43\textwidth]{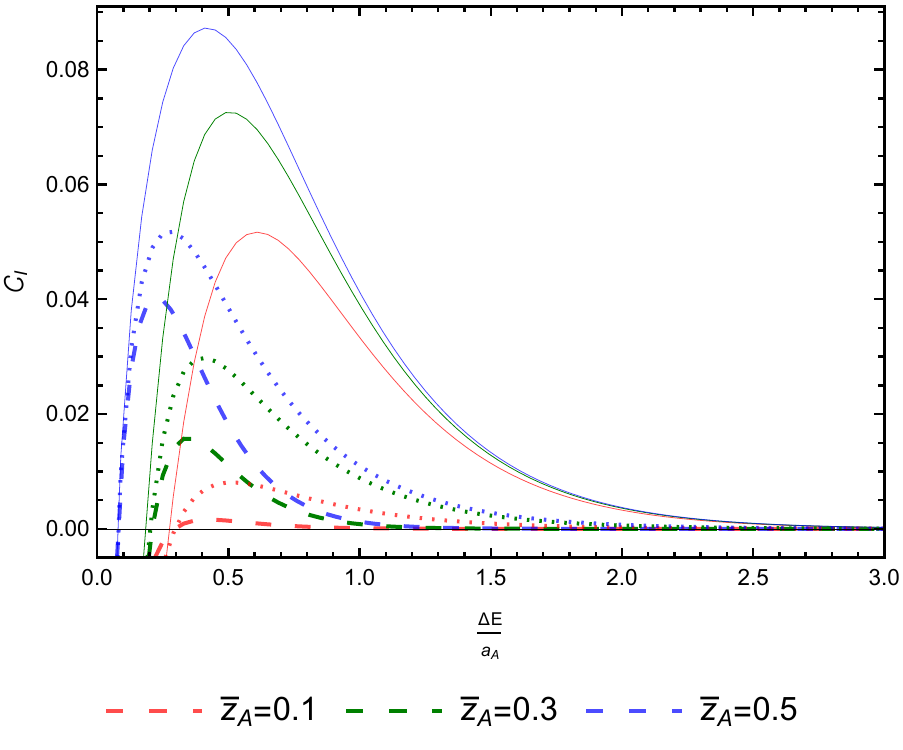}}{(a) $\bar{L}=1.0$}
\stackunder[5pt]{\includegraphics[width=0.4\textwidth]{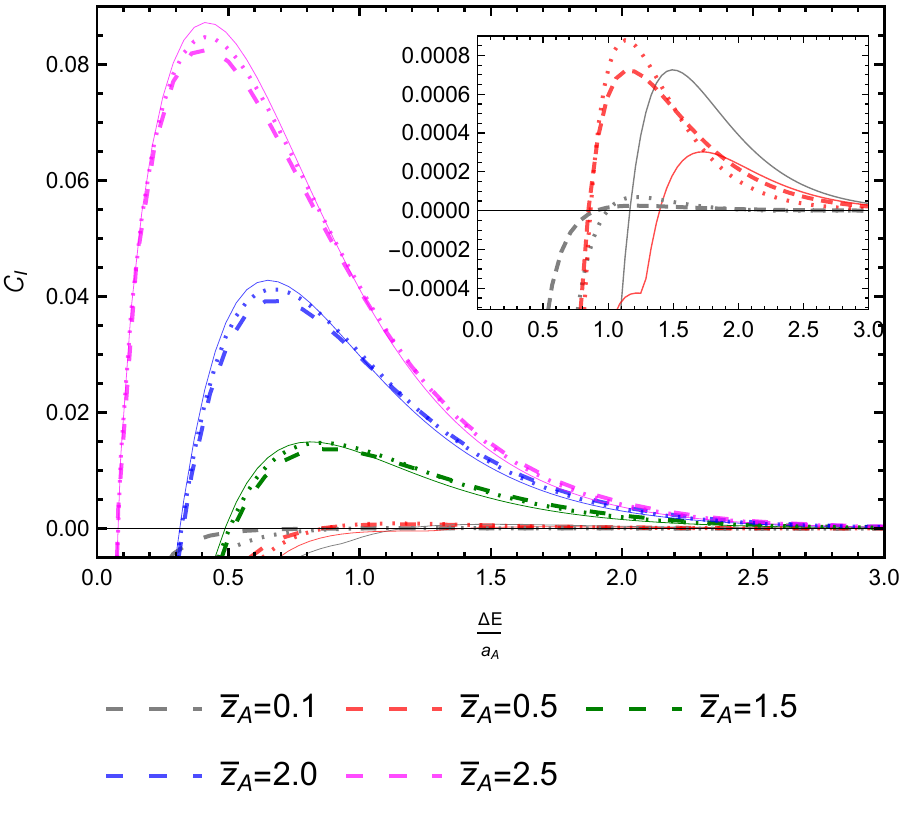}}{(b) $\bar{L}=5.0$}\\
\stackunder[5pt]{\includegraphics[width=0.43\textwidth]{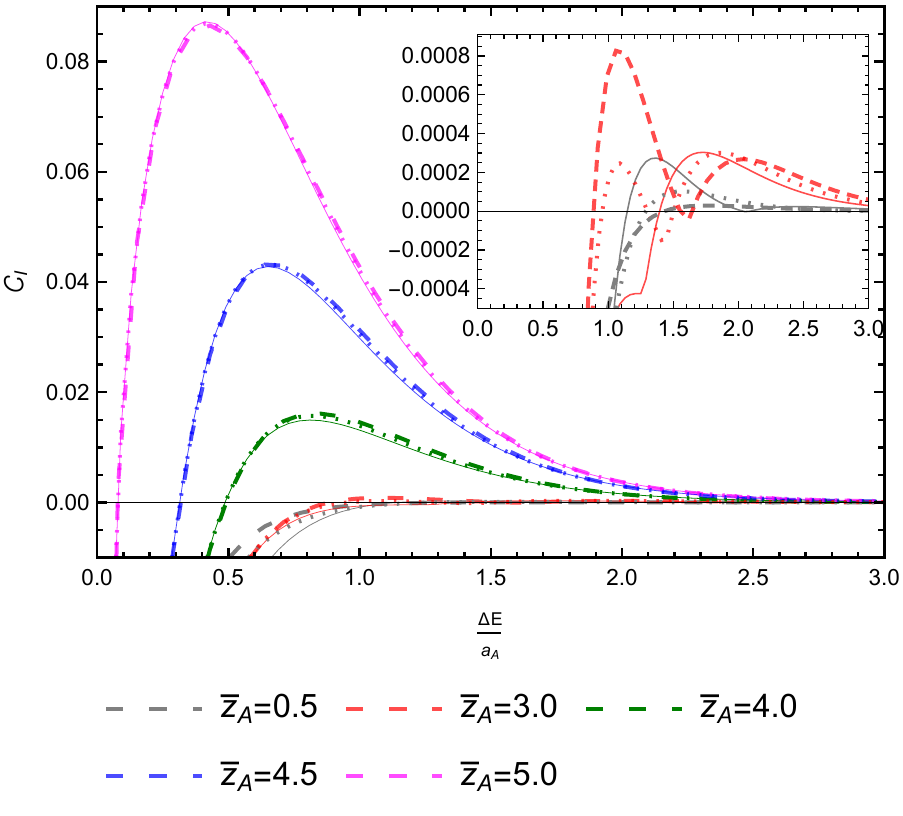}}{(c) $\bar{L}=10.0$}
\caption{We plotted $\mathcal{C}_{I}$ with respect to the dimensionless inverse acceleration $\Delta{E}/a_{A}$: (a) for $\bar{L}=1.0$, (b) for $\bar{L}=5.0$ and (c) for $\bar{L}=10.0$, respectively. Different colours are used for different fixed values of  $\bar{z}_{A}$ with the constraint $\bar{z}_{A}+\bar{z}_{B}=\bar{L}$. Here we used solid, dotted and dashed lines to represent no boundary, single boundary and double boundary systems, respectively.}
	\label{fig:figures1}
\end{figure}

 We consider that both detectors are accelerating in an anti-parallel manner along the $x$-direction. The detector $A$ is positioned near the boundary at $\bar{z}=0$, and the detector $B$ is near the boundary at $\bar{z}=\bar{L}$. Both detectors are equally distanced from the $\bar{z}=\bar{L}/2$ plane, which is implemented by the constraint $\bar{z}_{A}+\bar{z}_{B}=\bar{L}$. We also consider the same positions of the detectors for the no-boundary and single-boundary systems to compare the concurrence among them. Therefore we use the same constraint $\bar{z}_{B}=\bar{L}-\bar{z}_{A}$ in the expressions of $\mathcal{C}_{I}$ for the single and no boundary systems with $0<\bar{z}_{A}<\bar{L}$. Note that $\bar{L}=L_0\Delta{E}=L\Delta{E}$ for the double boundary system and for single and no boundary systems, $\bar{L}=L_0\Delta{E}$ is just a numerical parameter.

In Fig. \ref{fig:figures1}, we plot $\mathcal{C}_{I}$ with respect to the dimensionless inverse acceleration of the detector $A$ (i.e., $\Delta{E}/a_{A}$) with (a) $\bar{L}=1.0$, (b) $\bar{L}=5.0$ and (c) $\bar{L}=10.0$, respectively. We also choose different colours to describe the results with different $\bar{z}_{A}$ values. In these plots, one can see that entanglement harvesting is possible only in a particular range of acceleration values, depending on the other parameters $\bar{L}$, $\bar{z}_{A}$ and number of boundaries in the considered systems. 
For lower separation between the boundaries ($\bar{L}=1.0$) in subfigure \ref{fig:figures1}(a), we observe that $\mathcal{C}_{I}$ for any particular value of $\Delta{E}/a_{A}$ and $\bar{z}_{A}$, has maximum value for the no boundary system and minimum value for the two boundary system. Also, for a particular value of $\bar{z}_{A}$, the allowed range of acceleration for entanglement harvesting is much suppressed for the double boundary system and less suppressed for the single boundary system. However, as $\bar{L}$ increases, we can have different scenario. For instance, with $\bar{L}=5.0$ and $10.0$ (see, subfigures \ref{fig:figures1}(b) and \ref{fig:figures1}(c)), one observes that the allowed range of accelerations for entanglement harvesting in the single and double boundary systems are almost equal to that of the no boundary system. Also, suppression of the peak of $\mathcal{C}_{I}$ for any particular value of $\Delta{E}/a_{A}$ and $\bar{z}_{A}$ is very small for the single and double boundary systems compared to the no boundary system. For $\bar{L}=5.0$, subfigure \ref{fig:figures1}(b) shows that for any particular value of $\bar{z}_{A}$, the concurrence quantity has maximum suppression for the double boundary system for a smaller $\Delta{E}/a_{A}$ value. However, for a higher $\Delta{E}/a_{A}$ value and any fixed $\bar{z}_{A}$, there is enhancement in $\mathcal{C}_{I}$ quantity compared to the no boundary $\mathcal{C}_{I}$ quantity. The maximum enhancement is always for the double boundary system. The similar nature of enhancement in $\mathcal{C}_{I}$ is also observed for $\bar{L}=10.0$ (see, subfigure \ref{fig:figures1}(c)). Thus it appears that the presence of reflecting boundaries suppresses entanglement harvesting between two detectors for small $\bar{L}$ values. However, compared to an unbounded situation, the entanglement harvesting can be enhanced by introducing reflecting boundaries with a large separation between them.  

%%%%%%%%%%%%%%%%%%%%%%%%%%%%%%%%%%%%%%%%%%%%%%%%%%%%%%%%%%%%%%
\begin{figure}[h!]
	\centering
\includegraphics[width=0.53\textwidth]{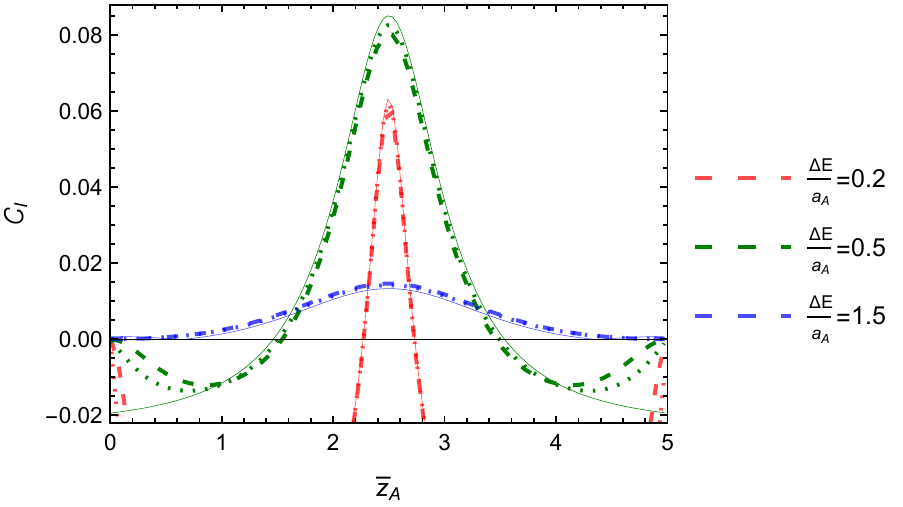}
\caption{We plotted $\mathcal{C}_{I}$ with respect to $\bar{z}_{A}$ with $\bar{L}=5.0$ and $\bar{z}_{A}+\bar{z}_{B}=\bar{L}$. Different colours are used for different fixed values of $\Delta{E}/a_{A}$. Here we used solid, dotted and dashed lines to represent no boundary, single boundary and double boundary systems, respectively.}
	\label{fig:figures2}
\end{figure}

It is also perceivable that for any particular $\bar{L}$, as the value of $\bar{z}_{A}$ goes from $0$ to $\bar{L}/2$, the distance between the detectors decreases. As a consequence, the concurrence quantity for a particular $\Delta{E}/a_{A}$ increase as $\bar{z}_{A}$ goes from $0$ to $\bar{L}/2$ (in allowed parameter range, where $\mathcal{C}_{I}>0$). After crossing the value of $\bar{L}/2$, for any $\bar{z}_{A}=\bar{L}/2+d\,\,(\leq\bar{L})$, the concurrence quantity at any particular $\Delta{E}/a_{A}$ will have the same value as it has for $\bar{z}_{A}=\bar{L}/2-d$ (see, Fig. \ref{fig:figures2}).
This symmetrical nature of $\mathcal{C}_{I}$ around $\bar{z}=\bar{L}/2$ is expected due to the symmetry ($\bar{z}_{A},\,\bar{z}_{B})=(\bar{z}_{B},\,\bar{z}_{A})$ in the Wightman function in Eq. (\ref{green+2}).

%%%%%%%%%%%%%%%%%%%%%%%%%%%%%%%%%%%%%%%%%%%%%%%%%%%%%%%%%%%%%%
\subsubsection{Case-II}
\begin{figure}[h!]
	\centering
	\footnotesize
\stackunder[5pt]{\includegraphics[width=0.43\textwidth]{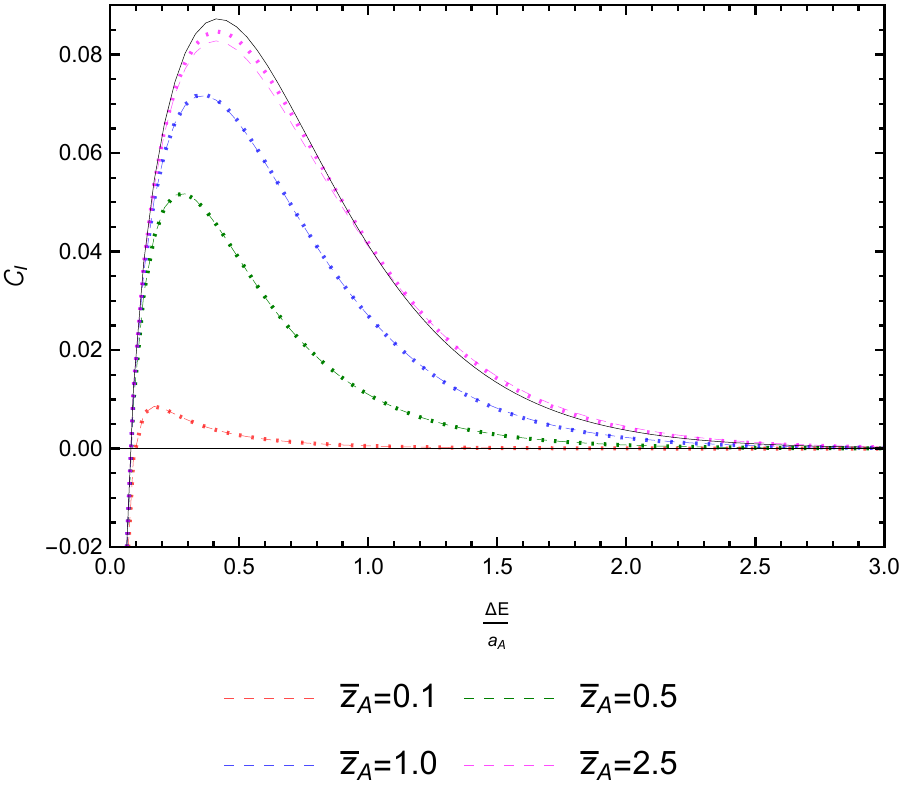}}{(a) $\bar{L}=5.0$, $\bar{z}_{A}=\bar{z}_{B}$}
\stackunder[5pt]{\includegraphics[width=0.43\textwidth]{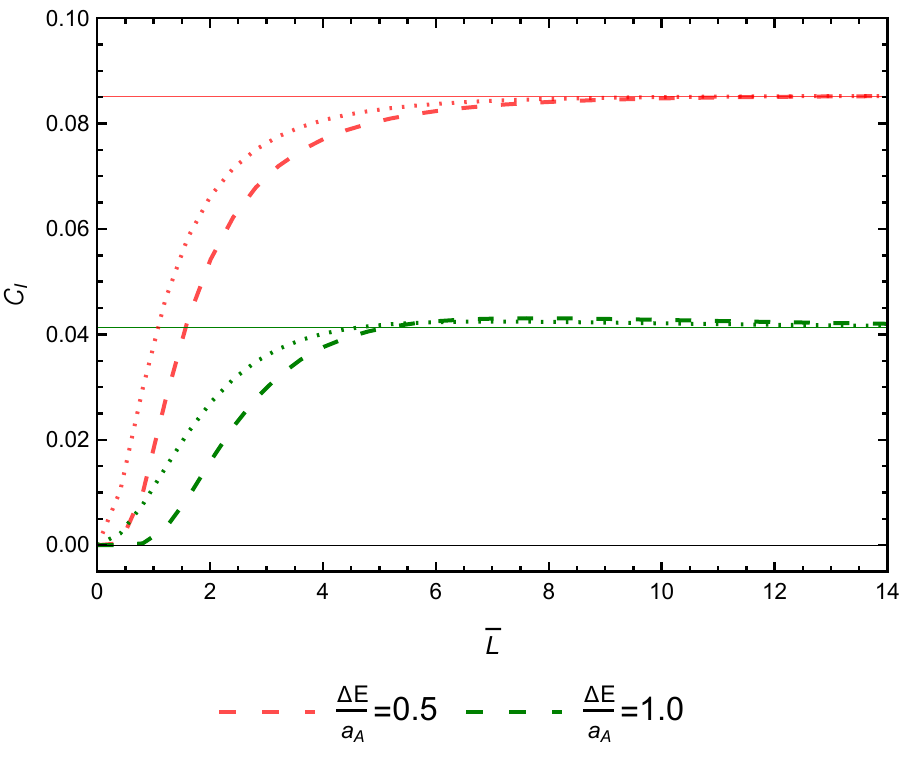}}{(b) $\bar{z}_{A}=\bar{z}_{B}=\bar{L}/2$}
\caption{(a) We plotted $\mathcal{C}_{I}$ with respect to $\Delta{E}/a_{A}$ with $\bar{L}=5.0$ and $\bar{z}_{A}=\bar{z}_{B}$. Different colours are used for different $\bar{z}_{A}$ values. (b) We plotted $\mathcal{C}_{I}$ with respect to $\bar{L}$ with consideration of $\bar{z}_{A}=\bar{z}_{B}=\bar{L}/2$. Different colours are used for different fixed $\Delta{E}/a_{A}$ values. Here we used solid, dotted and dashed lines to represent no boundary, single boundary and double boundary systems, respectively.}
	\label{fig:figures3}
\end{figure}

After analysing the case where the detectors have a different perpendicular separation between them, here we consider the situation where the detectors have a fixed perpendicular separation ($\bar{z}_{A}=\bar{z}_{B};\,\bar{\Delta}{y}=0.1$). Again we consider the detectors to be accelerating in an anti-parallel manner along the $x$-axis. Keeping the separation between the detectors fixed, we take the $\bar{z}$-coordinates of both detectors between $0$ and $\bar{L}/2$. Therefore, the change in $\mathcal{C}_{I}$ for different $\bar{z}_A=\bar{z}_B$ is solely due to the influence of the boundaries.

Therefore, for the no boundary system, $\mathcal{C}_{I}$ has the same value for a particular value of $\Delta{E}/a_{A}$ with any fixed $\bar{z}_{A}$. 
For smaller values of $\bar{L}$ (say, $\bar{L}=1.0$), $\mathcal{C}_{I}$ will be suppressed due to the presence of boundary. The suppression of the $\mathcal{C}_I$  quantity for the single and double boundary system is already observed for $\bar{z}_{A,B}=\bar{L}/2$ (with $\bar{L}=1$) in subfigure \ref{fig:figures1}(a). The suppression will be even higher for other $\bar{z}_{A,B}$ values when $\bar{z}_{A,B}<\bar{L}/2$. Note that, the double boundary $\mathcal{C}_{I}$ has symmetrical nature around $\bar{z}=\bar{L}/2$ for any particular value of $\Delta{E}/a_{A}$ (similar to the case-I). However, this is not true for the single boundary system as both detectors keep moving away from the boundary at $\bar{z}=0$. Further increasing $\bar{z}_{A}$, the single boundary $\mathcal{C}_{I}$ will eventually become the same for the no boundary system. 

However, the enhancement in the concurrence quantity is only possible for larger $\bar{L}$ values ($\bar{L}\gtrsim5.0$).
In subfigure \ref{fig:figures3}(a), we have shown this no boundary $\mathcal{C}_{I}$ quantity in a black solid line, while the single and double boundary systems are shown in dotted and dashed lines, respectively. Here we observe that as $\bar{z}_{A,B}$ increases, the concurrence for single and double boundary systems at any particular $\Delta{E}/a_{A}$ increase for any fixed value of $\bar{L}$. For fixed $\bar{z}_{A}=\bar{L}/2\,\,(=2.5)$ and a higher value of $\Delta{E}/a_{A}$, there is enhancement in concurrence quantity for the single and double boundary systems.  However, the latter one enjoys slightly more. Also, the allowed ranges of accelerations for entanglement harvesting increase with $\bar{z}_{A}$. 
In Fig. \ref{fig:figures3}(b), we have plotted the concurrence quantity $\mathcal{C}_{I}$ with respect to $\bar{L}$ with consideration that $\bar{z}_{A,B}=\bar{L}/2$ (remember for double boundary systems $\bar{L}=L_{0}\Delta{E}=L\Delta{E}$, therefore both positions of the detectors and the second boundary is changing). Here we again see that $\mathcal{C}_{I}$ for single and double boundary systems increase with $\bar{L}$ for a fixed value of $\Delta{E}/a_{A}$. Entanglement enhancement is observed for a large value of $\Delta{E}/a_{A}\,(=1.0)$, which is more for the double boundary system.

\subsubsection{Case-III}

\begin{figure}[h!]
	\centering
	\includegraphics[width=0.43\textwidth]{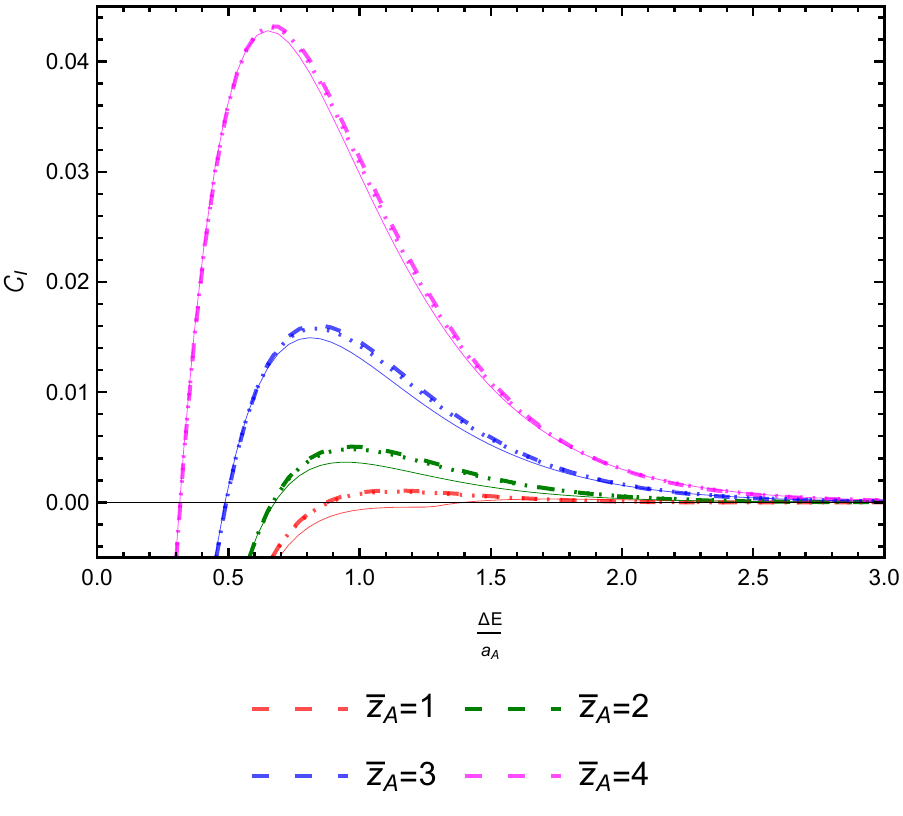}
\caption{We plotted $\mathcal{C}_{I}$ with respect to $\Delta{E}/a_{A}$ and fixed values of $\bar{z}_{A}$. Here we used $\bar{z}_{B}=5.0$ and $\bar{L}=10.0$. Different colours are used for different $\bar{z}_{A}$ values. Here we used solid, dotted and dashed lines to represent no boundary, single boundary and double boundary systems, respectively.}
	\label{fig:figures4}
\end{figure}
Finally, we consider a situation where detector $B$ is fixed at $\bar{z}=5.0\,(\bar{L}=10.0)$ and different $\bar{z}$-positions for detector $A$ has taken in the range of $0<\bar{z}_{A}<5.0$. Here again, we consider the detectors to be accelerating in an anti-parallel manner along the $x$-axis. We observe a boundary-induced enhancement in $\mathcal{C}_{I}$ for any $\Delta{E}/a_{A}$ value (in the allowed ranges of accelerations) with all fixed $\bar{z}_{A}$ values (see, Fig. \ref{fig:figures4}). We also see that $\mathcal{C}_{I}$ for double, single and no boundary systems increase as $\bar{z}_{A}$ is approaching $\bar{z}_{B}$ for any choice of $\bar{L}$. The entanglement amplification due to the double boundary is more perceptible compared to the single boundary system. Like the previous cases, enhancement in $\mathcal{C}_{I}$ due to the presence of boundary is only possible for larger $\bar{L}$ values, not for smaller $\bar{L}$ values.

%%%%%%%%%%%%%%%%%%%%%%%%%%%%%%%%%%%%%%%%%%%%%%%%%%%%%%%%%%%%%%%%%%%%%%%%%%%%

\section{Discussion and implications}\label{sec:5}

We have investigated the influence of multiple reflecting boundaries on entanglement harvesting between two uniformly accelerated UDW detectors. In literature, existing studies suggest that entanglement harvesting in the presence of a single reflecting boundary can get suppressed or enhanced depending on the parameter space. However, no studies have been conducted on whether increasing the number of reflecting boundaries enhances similar features. Here we have done a comparative study on entanglement phenomena between two detectors in the presence of double, single and no reflecting boundaries. We considered the monopole coupling model with the eternal switching function of the interaction to obtain a simple analytic expression of the concurrence quantity. Due to this choice of the switching function, we found that entanglement extraction from the field vacuum is only possible for the anti-parallel motion of the detectors. Since we considered identical detectors with the same energy gap, their acceleration must have the same magnitude. We observe that detectors' entanglement increases as the vertical separation between them decreases for any number of boundaries. For the single and double boundary systems, the entanglement gets suppressed if any one or both of the detectors are near the boundary or boundaries. Entanglement degradation is much higher for the double boundary system than the single boundary system. Entanglement harvesting increases as the detectors move away from the boundary or boundaries. For small separations between the boundaries, the influence of the boundaries is strong, leading to higher degradation. As the separation increases, the boundary influence on the detectors decreases; the concurrence approaches the same for no boundary system. In some specific parameter spaces, the double boundary concurrence crosses the free space as well as the single boundary situations. Similar nature of concurrence is also found for the single boundary system, where the degradation and the enhancement of the entanglement only depend on distance from the first boundary.
One of the important observations is -- the double boundary concurrence degrades more whenever there is a degradation. The same also holds for the enhancement of entanglement harvesting. Therefore an overall conclusion can be drawn that the presence of a more number of reflecting boundaries enhances the similar effect observed for a single reflecting boundary system.

\vskip 4mm
\noindent
{\bf Acknowledgments:}
DB would like to acknowledge Ministry of Education, Government of India for providing financial support for his research via the Prime Minister's Research Fellows (PMRF) May 2021 scheme. The research of BRM is supported by a START-UP RESEARCH 
GRANT (No. SG/PHY/P/BRM/01) from the Indian Institute of Technology Guwahati, 
India.

\begin{appendix}
\section*{Appendix}
%{\Large \textbf{Appendix}}
\section{Finite summation for Eq. (\ref{Ent2})}\label{AppenA}

\begin{figure}[h!]
	\centering
	\footnotesize
\stackunder[5pt]{\includegraphics[width=0.43\textwidth]{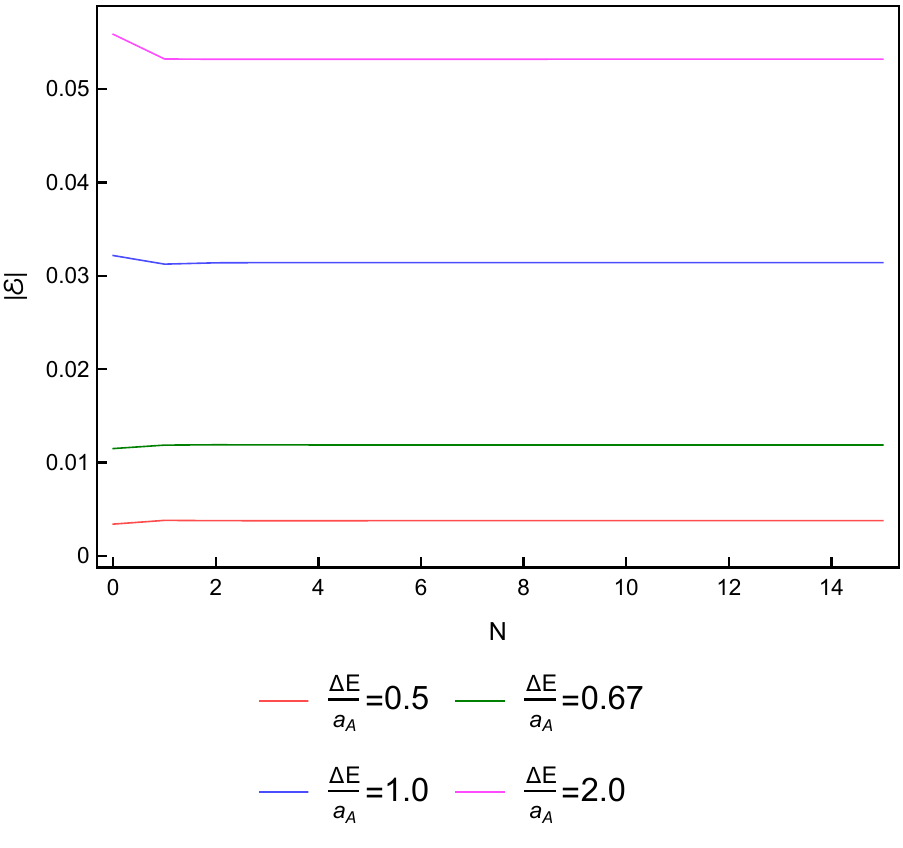}}{(a) $0\leq N\leq15$}
\stackunder[5pt]{\includegraphics[width=0.445\textwidth]{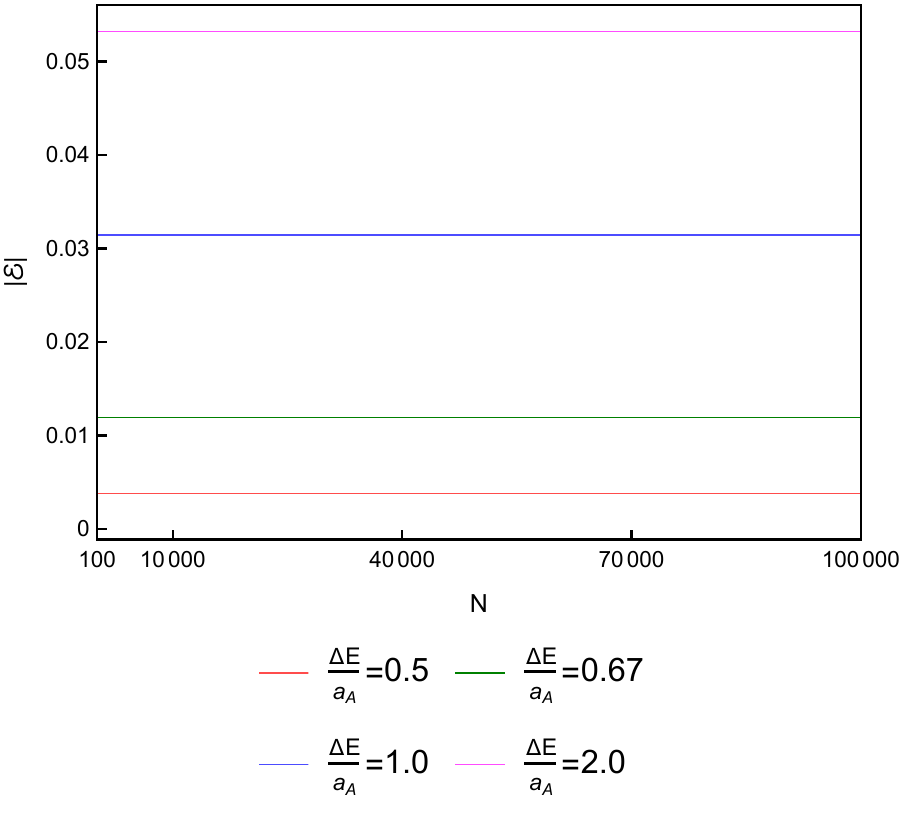}}{(b) $100\leq N\leq100000$}
\caption{In subfigures (a) and (b), we plotted $|\mathcal{E}|$ with respect to $N$ for different $\Delta{E}/a_{A}$ values. 
Here we used $\bar{\Delta{y}}=0.1,\,\, \bar{z}_{A}=2.0,\,\, \bar{z}_{B}=3.0$ and $\bar{L}=5.0$. Different colours are used for different $\Delta{E}/a_{A}$ values.}
	\label{fig:figures5}
\end{figure}

After Eq. (\ref{Ent2}), we analytically argued why one can perform a finite sum over $n$ instead of the infinite sum. Here we give two plots in Fig. \ref{fig:figures5}, which show how the absolute value of the entangling term changes with respect to $\max\{n\}=N$ with other parameters are fixed. Here we consider $\bar{\Delta{y}}=0.1,\,\, \bar{z}_{A}=2.0,\,\, \bar{z}_{B}=3.0$, $\bar{L}=5.0$ and $\Delta{E}/a_{A}=0.5,\,0.67,\,1.0,\,2.0$. We used different colours to represent different $\Delta{E}/a_{A}$ values. These plots show that the entangling term changes for small values of $N$, while it remains constant for large $N$.
Note that for $N=0$, the quantity $|\mathcal{E}|$ corresponds to the single boundary system.  

The same is also true for the other fixed parameter values.

%%%%%%%%%%%%%%%%%%%%%%%%%%%%%%%%%%%%%%%%%%%%%%%%%%%%%%%%%%%%%%%%%%%%%%%%%%%%%%%%%%%%%%%%%%%%%%%%%%%%%%%%%%%%%%%%%%%%%%%%%%%%%%%%%%
\iffalse
\begin{table}[h]
\begin{center}
\caption{Values of $|\mathcal{E}|$ obtained for different values of $N$ for $\Delta{E}/a_{A}=0.5,\,0.67,\,1.0,\,2.0$. Other parameter values are $\bar{\Delta{y}}=0.1,\,\, \bar{z}_{A}=1.0,\,\, \bar{z}_{B}=2.5$ and $\bar{L}=5.0$. }
\label{table:ep}
\begin{tabular}{|c | c| c| c|c|}
\hline
$N$&$\Delta{E}/a_{A}=0.5$ &$\Delta{E}/a_{A}=0.67$& $\Delta{E}/a_{A}=1.0$ &  $\Delta{E}/a_{A}=2.0$\\
\hline
0&0.0023299725541837&0.0075602197807939&0.019760109923661&0.029390035717016
\\
1&0.0025614008300671&0.0078666436655535&0.019647862989064&0.028561295240237
\\
2&0.0025468459207773&0.0078797191135678&0.019704250934533&0.028548021614454
\\
5&0.0025455260834315&0.0078707304256613&0.019713522569669&0.0285516167466
\\
10&0.002545723794244&0.0078709957968478&0.019712492673788&0.028552699015268
\\
100&0.002545676374149531&0.00787108571010797&0.01971232769294962&0.028552905278051467
\\
500&0.002545676438030358&0.00787108575283721&0.019712327581865106&0.028552905098059092
\\
1000&0.002545676438598018&0.007871085754018224&0.019712327582645423&0.02855290509684143
\\
10000&0.002545676438531245&0.007871085753975254&0.01971232758287441&0.02855290509684507
\\
20000&0.0025456764385313226&0.007871085753975132&0.01971232758287425&0.0285529050968452
\\
100000&0.002545676438529336&0.007871085753961704&0.01971232758286726&0.028552905096815046
\\
\hline
\end{tabular}
\end{center}
\end{table}
\fi
\end{appendix}

  \bibliography{bibtexfile}
\bibliographystyle{ieeetr}
\newpage
\end{document}